\documentstyle[aps,prl,floats]{revtex}

\begin{document}

\def\simg{\mathrel{\hbox{\rlap{\lower.55ex \hbox {$\sim$}}
                   \kern-.3em \raise.4ex \hbox{$>$}}}}
\def\siml{\mathrel{\hbox{\rlap{\lower.55ex \hbox {$\sim$}}
                   \kern-.3em \raise.4ex \hbox{$<$}}}}
\def\bitm{\bibitem}
\def\beq{\begin{equation}}
\def\enq{\end{equation}}
\def\bea{\begin{eqnarray}}
\def\ena{\end{eqnarray}}
\def\nonum{\nonumber}
\def\bec{\begin{center}}
\def\enc{\end{center}}
\def\etal{{\it et al.}}
\def\to{\rightarrow}
\def\Mesz{M\'esz\'aros~}
\def\Pacz{Paczy\'nski~}
\def\barnu{\bar\nu}
\def\barnue{\bar \nu_e}
\def\barnum{\bar \nu_\mu}
\def\epm{\hbox{e}^\pm}
\def\nue{\nu_e}
\def\num{\nu_\mu}
\def\E53{E_{53}}
\def\L51{L_{52}}
\def\r13{r_{13}}
\def\cmcui{\hbox{cm}^{-3}}
\def\cmsqi{\hbox{cm}^{-2}}
\def\si{\hbox{s}^{-1}}
\def\msun{M_\odot}
\def\eps{\epsilon}
\def\epstev{\epsilon_{\nu,\rm TeV}}
\def\Tho{\Theta_o}
\def\Th{\Theta}
\def\st{\sigma_T}

\draft
\tighten
\twocolumn[\hsize\textwidth\columnwidth\hsize\csname@twocolumnfalse%
\endcsname

\title{TeV Neutrinos from Successful and Choked Gamma-Ray Bursts}

\author{Peter \Mesz$^{1}$ and Eli Waxman$^{2}$\\
%EndName
$^1$ Pennsylvania State University \\
$^2$ Weizman Institute of Science }
%hep-ph/0004019}

\maketitle

\begin{center} { {\it Phys.Rev.Lett}, {\footnotesize in press, accepted} 8/31/01 (astro-ph/0103275)} \end{center}

\begin{abstract}

Core collapse of massive stars resulting in a relativistic fireball jet
which breaks through the stellar envelope is a widely discussed scenario for 
$\gamma$-ray burst production. For very extended or slow rotating stars, 
the jet may be unable to break through the envelope. Both penetrating 
and choked jets will produce, by photo-meson interactions of accelerated 
protons, a burst of $\gtrsim5$~TeV neutrinos while propagating in the 
envelope. The predicted flux, from both penetrating and choked jets, 
should be easily detectable by planned 1~km$^3$ neutrino telescopes.

\end{abstract}

]

\narrowtext

%\section{Introduction}

The leading model for gamma-ray bursts (GRB) involves a relativistic 
fireball, where the $\gamma$-rays are produced by synchrotron or inverse 
Compton radiation from Fermi accelerated electrons in optically thin shocks. 
The ultimate energy source of the fireball is thought to be the gravitational 
energy release associated with temporary mass accretion onto 
a black hole, resulting either from a compact merger \cite{pac86} or 
from the collapse of a massive star \cite{woo93} (see \cite{m01} for 
a review).  All the GRB so far localized with the Beppo SAX satellite
\cite{cos99} belong to the class of long bursts \cite{kou93} with
durations $\Delta t_\gamma \simg 2$ s, and there is increasing evidence 
that these occur in star-forming galaxies \cite{bloom00}. 
For this class at least, comprising $\sim 2/3$ of all bursts, it is widely 
assumed that a massive, collapsing star (``collapsar") is the progenitor. 
The shocks producing the $\gamma$-rays must occur after the fireball has 
emerged from the stellar envelope. The preferred escape route is along 
the centrifugally lightened rotation axis, and the stellar pressure tends 
to collimate the fireball into a jet. The shocks should also lead to Fermi 
accelerated relativistic protons \cite{wax95,vie95}, and to neutrinos with 
$\eps_\nu\simg 10^{14}$ eV via interactions with the $\sim$ MeV $\gamma$-rays 
\cite{wb97,rm98} (see \cite{w01} for a review). 

% CH3
In this {\it Letter} we show that while a jet is making its way through the star,
its rate of advance is slowed down in a termination shock to a Lorentz factor 
much lower than it has at injection or after emergence; and additional internal 
shocks are expected in the jet interior, inward of the termination shock. 
% ECH3
The latter can accelerate protons to $\gtrsim 10^5$ GeV, which interact with 
X-ray photons in the jet cavity leading to electron and muon neutrinos 
(and anti neutrinos) with energies $\eps_{\nu} \gtrsim5$ TeV. These neutrinos 
appear as a precursor signal, lasting for time scales $\gtrsim$ tens of seconds 
prior to the observation of any $\gamma$-rays produced outside the star by a 
collapsar-induced GRB.

% CH ref A.1
Furthermore, in a significant fraction of the massive stellar collapses, the 
jet may be unable to punch through the stellar envelope \cite{macfadyen00}. 
However, the TeV neutrino signal from such choked jets should be similar to 
that from ``successful'' jets which do break through and lead to observable 
GRB. This may provide a means of detecting and counting such choked-off, 
$\gamma$-ray dark collapses.

The TeV fluence from an individual collapse at cosmological distance 
$z\sim 1$ implies $10^{-1}-10$ upward moving muons per collapse/burst in 
a 1~km$^3$ detector. 
The signal from an individual collapse with associated electromagnetic 
signals (and possibly from rare, energetic dark collapses), are
well above the atmospheric neutrino background, and the total integrated 
flux from dark collapses should also be detectable.
The AMANDA experiment \cite{halzen01,amanda-bound} may soon be able to 
provide relevant limits on the total integrated rate of such dark collapses.

\paragraph*{Jet propagation inside a massive progenitor envelope.}

Numerical simulations of collapsar models leading to black hole formation 
\cite{macfadyen00} have been performed for a range of stellar progenitors 
and initial conditions. The characteristic progenitor structure is that of 
an evolved massive star, with $\sim2M_\odot$ Fe core of radius $\sim 10^9$~cm 
and $\sim 8M_\odot$ He core extending to $\sim10^{11}$~cm, in some cases 
surrounded by an H envelope extending to $\gtrsim 10^{13}$~cm (a red 
supergiant), while in others the envelope has been (largely) lost. 
% CH3
While the details remain uncertain, preliminary calculations suggest 
that a relativistic jet can be launched along the progenitor rotation 
axis \cite{macfadyen00,aloy00}, powered either by thermal $\nu\barnu$ 
% ECH3
annihilation or MHD stresses coupling to the black hole and to the debris 
disk falling back onto the hole. The jet life time is limited by the gas 
fall-back time onto the black hole, and should be comparable to GRB durations, 
$\siml 10^2$~s. (Similar pair or MHD outflows may also arise from collapses 
leading to highly magnetized pulsars \cite{cen98,wheeler00}).

When $\gamma$-rays are observed, the required isotropic equivalent 
luminosities are $L_{\rm iso.} \gtrsim10^{52} {\rm erg/s}$ and the inferred 
jet Lorentz factors are $\Gamma_j \sim 10^2$. This is the final value
outside the star, which may be representative of the intrinsic injection
Lorentz factor. As it advances through the star, the jet drives a bow shock 
ahead of it. The jet is capped by a termination shock, and a reverse 
shock moves back into the jet, at which the relativistic jet plasma 
is decelerated. Both the shocked jet plasma and the shocked stellar plasma 
advance with a jet head Lorentz factor $\Gamma_h\ll\Gamma_j$ \cite{mr01}. 
In the rest frame of the shocked, decelerated plasma, the fast jet moves 
with a Lorentz factor $\approx\Gamma_j/2\Gamma_h$. During propagation in the 
Fe and He cores, the high density ahead of the jet implies deceleration of jet 
plasma to subrelativistic velocity. However, beyond the He core edge at 
$r\sim10^{11}$~cm, the density in an H envelope drops to $\rho\sim10^{-7} 
{\rm g\,cm}^{-3}$, and the jet can accelerate to relativistic velocity 
% CH3 
\cite{mr01} (c.f. also \cite{mckee01}.  Balancing the energy density behind 
the forward shock, $4\Gamma_h^2 \rho c^2$, with that behind the reverse shock, 
$4(\Gamma_j/2\Gamma_h)^2 n_j m_p c^2$, where the jet proper proton 
number density $n_j = L_{\rm iso}/(4\pi r^2\Gamma_j^2 m_p c^3)$,
%$n_j$ is given by $L_{\rm iso}=4\pi r^2\Gamma_j^2 n_j m_p c^3$, 
we find 
\begin{equation}
\Gamma_h=3\frac{L_{52}^{1/4}}{r_{12}^{1/2}\rho_{-7}^{1/4}},
\label{eq:Gamma_h}
\end{equation}
where $L_{\rm iso}=10^{52}L_{52}{\rm erg/s}$, $r=10^{12}r_{12}$~cm is the 
shock radius, and $\rho=10^{-7}\rho_{-7}{\rm g\,cm}^{-3}$. 

The jet head
propagates at mildly relativistic velocity $\Gamma_h$ up to a typical 
super-giant H envelope radius $r\sim10^{13}$~cm (and if it makes it through
entirely, in the even lower density beyond it would increase back to 
$\Gamma_j\sim 10^2\Gamma_2$).
%CH ref A.5
Envelopes larger than $R_\ast\gtrsim 10^{13}R_{13}$ cm, 
or denser cores, e.g. due to slower rotation rate, 
could however lead to a stalling of the jet before it breaks through 
the star, if the jet crossing time is longer than the jet lifetime 
(as argued by, e.g. \cite{macfadyen00,mr01}).

The proper number density $n$ of protons and electrons in the shocked, 
decelerated jet plasma is related to the proper density $n_j$ in the 
fast moving jet through $n=4(\Gamma_j/2\Gamma_h)n_j$, 
and the energy density in the shocked plasma is 
$U=4(\Gamma_j/2\Gamma_h)^2 n_j m_p c^2$.
The hot, shocked jet plasma drives the bow shock wave that 
propagates forward (radially) and sideways (tangentially) into the 
envelope. For a jet of small opening angle $\theta<1/\Gamma_h$ 
\cite{mr01}, the proper thickness of the shocked jet plasma shell is 
$\Delta\sim\theta r$. This can be seen by balancing the particle flux 
across the reverse shock, 
%standoff 
$\approx\pi(\theta r)^2c(\Gamma_j/2\Gamma_h) n_j$, 
with the tangential flux of particles leaving the cylinder of height 
$\Delta$ and radius $\theta r$ of shocked plasma, 
$\approx2\pi\theta r\Delta(c/\sqrt{3})4(\Gamma_j/2\Gamma_h) n_j$ 
(taking the tangential flow velocity to be comparable to the 
sound speed in the shocked plasma). Adopting $\Delta=0.2\theta r$  
and $\theta=10^{-1}\theta_{-1}$, the Thomson optical depth of the 
shocked plasma shell is 
\begin{equation}
\tau_T\approx4\times10^3\frac{\theta_{-1}L_{52}}{r_{12}\Gamma_{2}\Gamma_h},
\label{eq:tau}
\end{equation}
where $\Gamma_j=10^2\Gamma_2$.

The reverse shock may start out being collisionless, heating the electrons and 
protons to $\sim 10^2\Gamma_2$ GeV. However, the electrons would lose all 
their energy on very short time scale, by synchrotron and inverse-Compton 
emission, converting a large fraction of the shocked plasma internal energy 
$U$ to radiation. Due to the high optical depth, the radiation will thermalize, 
with an approximate black body radiation temperature 
\begin{equation}
T_r\approx 4 \left(\frac{L_{52}}{r_{12}^2 \Gamma_h^2}\right)^{1/4}\,{\rm keV}.
\label{eq:T_r}
\end{equation}
The corresponding proper photon number density is 
\begin{equation}
n_\gamma\approx 2\times10^{24} 
\left({L_{52}\over r_{12}^2 \Gamma_h^2}\right)^{3/4}\,{\rm cm}^{-3}.
\label{eq:n_g}
\end{equation}
Thus, the reverse shock is likely to become radiation 
dominated, i.e. the dissipation of jet kinetic energy may be mediated by 
inverse-Compton scattering of the electrons, whose
mean-free-path is $\sim1/n_\gamma\sigma_T\approx1$~cm, where $\sigma_T$ is
the Thomson cross section, and electrons lose all their momentum in a small 
number of scatterings (scattering of a thermal photon boosts its energy to 
$\sim 10 (\Gamma_2/\Gamma_h)^2$~MeV in the shocked plasma frame). 
%These high energy photons would produce several generations of 
%$\epm$ pairs, degrade their energy and then thermalize. Thus the 
%shock may also be mediated by radiative dissipation, both the typical 
%gyroradius and the radiative dissipation mean free path being $\sim 1$ cm.

The mean-free-path of thermal photons propagating back into the jet,
in the decelerated plasma frame, is
\begin{equation}
l_\gamma\approx \Gamma_h/(2\Gamma_j n_j\sigma_T)=5\times10^{6} 
\frac{\Gamma_2\Gamma_h r_{12}^2}{L_{52}}\,{\rm cm}.
\label{eq:l_g}
\end{equation}
Thus, the thermal photons do not penetrate much into the jet. If the reverse 
shock is indeed radiation dominated, the shock thickness would also be 
of order $l_\gamma$.

\paragraph*{Photo-meson Interactions and Neutrinos.}

In addition to the bow and reverse shocks moving with $\Gamma_h \ll \Gamma_j$
in the star, internal shocks in the pre-deceleration jet moving with 
$\Gamma_j$ are expected to occur at radii 
\begin{equation}
r_s\approx\Gamma_j^2 c\delta t=3\times10^{11}\Gamma_2^2\delta t_{-3}\,{\rm cm},
\label{eq:r_s}
\end{equation}
which are smaller than the termination and reverse shock radius.
Here $\delta t$ is the variability timescale of the injected relativistic 
outflow, whose minimum value s $\sim 10^{-3} \delta t_{-3}$ s \cite{m01,w01}.
%is the light crossing time of the last stable orbit (at 3 Schwarzschild radii) 
%around a non-rotating $10M_\odot$ black hole.
% CH3 
While the details could depend on the ratio of MHD to pair and baryon energy,
it is commonly assumed that internal shocks are mildly relativistic 
% ECH3
in the jet frame, and protons will be Fermi accelerated to a power law distribution 
which approximates $dn_p/d\eps_p\propto \eps_p^{-2}$ (see \cite{w01} for 
detailed discussion).

As they approach the reverse shock, high energy protons may produce 
pions in photo-meson interactions with thermal photons of energy 
$\eps_\gamma \sim T_r \sim 4$ keV, 
provided their energy (measured in the decelerated plasma frame)
satisfies $\eps_p \eps_\gamma \gtrsim 0.3 {\rm GeV}^2$, i.e. for 
$\eps_p\gtrsim10^5$~GeV. 
The maximum energy to which protons may be accelerated in the internal
shocks is limited by synchrotron losses. The acceleration time, in
the jet frame, to observed energy $\epsilon'_p=\Gamma_h\epsilon_p$,
corresponding to energy $\approx\epsilon_p$ in the decelerated plasma frame,
is approximately \cite{wax95,w01}
\beq
t_a\sim \frac{R_L}{c}= \frac{\epsilon'_p}{\Gamma_j eB}=  
        10^{-6.5}\frac{\Gamma_2^2\Gamma_h
        \delta t_{-3}}{(L_{52}\xi_{B,-2})^{1/2}}
          \epsilon_{p,15}{\rm\,s}, 
\label{eq:t_a}
\enq
% CH B 6.b 
while the (jet frame) time scale for synchrotron losses, which dominate 
at high energy, is 
\begin{equation}
t_{\rm syn}=\frac{6\pi m_p^4 c^3\Gamma_j}{\sigma_T m_e^2\epsilon'_p B^2}=
10^{2}\frac{\Gamma_2^7\delta t_{-3}^2}
{\Gamma_h L_{52}\xi_{B,-2}\epsilon_{p,15}}{\rm\,s}.
\label{eq:t_syn}
\end{equation}
Here, $\epsilon_{p}=10^{15}\epsilon_{p,15}$~eV,
$R_L$ is the Larmor radius and 
$\xi_B=10^{-2}\xi_{B,-2}$ is the magnetic field equipartition fraction,
$4\pi r_s^2 c\Gamma_j^2 B^2/8\pi=\xi_B L_{\rm iso.}$.
Thus, protons may be accelerated to energies well above the
pion threshold.

Protons at the threshold energy, for which the thermal photons are at the 
$\Delta$-resonance, lose $\approx 20\%$ of their energy in each interaction. 
Inside the jet, the thermal photon density drops exponentially with the 
distance $x$ from the reverse shock, $n_\gamma(x)\propto\exp(-x/l_\gamma)$, 
and resonant protons will lose all their energy to pion production at a 
distance $x$ from the termination shock where the photon density satisfies 
\begin{equation}
n_\gamma^\pi\approx 5/(l_\gamma \sigma_{p\gamma})=
2\times10^{21}\frac{L_{52}}{\Gamma_2 \Gamma_h r_{12}^2}{\rm cm}^{-3}.
\label{eq:n_gpi}
\end{equation}
Here, $\sigma_{p\gamma}=5\times10^{-28}{\rm cm^2}$ is the cross section at 
resonance. Typically, resonant protons and neutrons (which are produced with 
$\pi^+$'s in proton photo-pion interactions)
will lose their energy at $x/l_\gamma\approx10$, 
while higher energy nucleons, for which $\sigma_{p\gamma}$
is lower, will lose their energy to pions at shorter distance.

The pions produced in photo-meson interactions lose energy by
inverse-Compton scattering on thermal photons. The lowest energy pions produced
are characterized by Lorentz factors comparable to those of the resonant
protons, $\approx10^5(T_r/4{\rm keV})^{-1}$ in the decelerated plasma frame. 
Inverse-Compton scattering of $\sim4$~keV photons by pions of higher 
energy is in the Klein-Nishina regime. Thus,
the inverse-Compton energy loss time of high energy pions is
$\tau_{IC}\approx[3n_\gamma^\pi\sigma_T(m_e/m_\pi)^2c/8]^{-1}
(\eps_\pi/10^{13}{\rm eV})(T_r/4{\rm keV})$. 
The ratio of $\tau_{IC}$ to the pion life time is
\begin{equation}
\frac{\tau_{IC}}{\tau_{\rm decay}}\approx5
\Gamma_2\Gamma_h^{1/2}\left(\frac{r_{12}^{2}}{L_{52}}\right)^{3/4}.
\label{eq:tpi}
\end{equation}
%$ \left({\tau_{IC}}/{\tau_{\rm decay}}\right) \approx3\times10^3
%\Gamma_2(r_{13}^{2}t_2/E_{53})^{3/4}$.
We therefore expect $\approx1/4$ of the energy lost by protons to pion 
production to be converted to muon neutrinos ($\approx1/2$ of the energy 
is lost to neutral pion production, and $\approx1/2$ of the charged pion 
energy is converted by decay into muon neutrinos).

If the termination shock is collisionless, rather than radiative, protons 
may be accelerated to energies exceeding the photo-meson threshold also in 
the termination shock. However, it is clear form Eq. (\ref{eq:tpi}) that 
%In this case, proton acceleration is limited not only 
%by synchrotron, but also by inverse-Compton losses from scattering off 
%thermal photons, whose density is much higher than in the internal shocks. 
%(For protons above the photo-meson threshold, pion production losses will 
%dominate. However, we are interested in proton acceleration up to the 
%threshold energy). The inverse-Compton loss time, in the jet frame, due to 
%scattering of thermal photons with proper density (in the decelerated 
%plasma frame) given by Eq. (\ref{eq:n_g}), is 
%$t_{IC}\approx(3/4)m_p^4 c^3\Gamma/
%(\sigma_T m_e^2 2\Gamma_h\epsilon_p\Gamma^2 aT_r^4)
%\approx4\times10^{-8}\Gamma_h r_{12}^2/(L_{52}\Gamma_2 \epsilon_{p,15})$~s, 
%similar to the acceleration time to this energy. However, 
inverse-Compton losses of the pions in the high density photon field prevent 
the production of very high energy neutrinos from protons accelerated in 
the termination shock. 

The $\num(\barnum,\nu_e,\bar\nu_e)$ from pion decays due to protons 
accelerated in internal shocks will have a typical energy $5-10\% \eps_p$. 
% CH ref A.7
We therefore expect a flat power per decade neutrino spectrum \cite{wb97},
$\eps_\nu^2dn_\nu/d\eps_\nu\propto\eps_\nu^0$, at
\beq
\eps_{\nu}\gtrsim 2~\left(\frac{1+z}{2}\right)^{-1}
\frac{\Gamma_h^{3/2}r_{12}^{1/2}}{L_{52}^{1/4}}\,{\rm TeV}~.
\label{eq:epsnu}
\enq
% CH : Main comment of ref B
We note, that since photons propagating backward into the jet are emitted from
a region of shocked, decelerated plasma with small optical depth, the
spectrum of photons with which protons interact may differ from
thermal. In particular, a non-thermal extension of the 
photon spectrum to energies
$\gg T_r$ may lead to an extension of the neutrino spectrum to energies
$<1$~TeV. A detailed calculation of the photon spectrum is, however, beyond
the scope of this {\it Letter}.

% CH ref A.8 
% CH ref A.9
The jet isotropic-equivalent luminosity, $L_{\rm iso.}\sim10^{52}{\rm erg/s}$,
and energy, $E_{\rm iso.}\sim10^{53}$~erg, are inferred from $\gamma$-ray 
observations, and reflect the energy of electrons accelerated to high energy
by the internal, mildly-relativistic collisionless shocks. 
% CH2
Protons are generally expected to be accelerated in such shocks 
with higher efficiency (see e.g. \cite{Blandford87} for review). 
We therefore assume that the accelerated proton energy is at least 
comparable to that inferred from observations to be carried
by accelerated electrons (see also \cite{wax95,vie95,wb97,alvarez00}).
A large fraction of this accelerated proton energy 
will be converted, while the jet is inside the envelope, to pions. 
The envelope is transparent to TeV neutrinos, and the 
expected TeV neutrino fluence at Earth from a jet at a 
luminosity distance $D_L\sim 10^{28}$ cm (redshift $z\sim1$) is
\beq
F_{\nu_\mu(\bar\nu_\mu,\nu_e,\bar\nu_e}
\sim\frac{E/8}{4\pi D_L^2}=10^{-5}\frac{E_{53}}{D_{28}^{2}}
{\rm erg/cm}^2,
\label{eq:fnu}
\enq
where $E_{\rm iso.}=10^{53}E_{53}$~erg.

The probability that a muon neutrino will produce a high-energy upward 
moving muon in a terrestrial detector is approximately given by the ratio 
of the muon range to the neutrino mean free path \cite{gaisser-rev},
$P_{\nu\mu}\approx1.3\times10^{-6} \epstev^{\beta}$, with $\beta=2$
for $\epstev <1$ and $\beta=1$ for $\epstev >1$. 
For a flat neutrino spectrum, this implies that the number flux $J_\mu$ 
of muon induced neutrinos can be related to the neutrino energy flux 
$F_\nu$ by $J_\mu=(P_0/\epsilon_{\nu,0})F_\nu$,
with $P_0/\epsilon_{\nu,o}=1.3\times10^{-6}{\rm TeV}^{-1}$. Using Eq. 
(\ref{eq:fnu}), the average number of upward muon 
detections for an individual burst is 
\begin{equation}
N_\mu\approx\frac{P_0}{E_0}\frac{E/4}{4\pi D_L^2}=0.2
\frac{E_{53}}{D_{28}^{2}}{\rm km}^{-2}\,.
\label{eq:N_mu}
\end{equation}

\paragraph*{Implications.}

The typical angular resolution of planned neutrino telescopes at TeV 
energies (see \cite{halzen01} for a review) should be $\theta\sim 1$ degree. 
The atmospheric neutrino background flux is 
$\Phi_{\nu,bkg} \sim 10^{-7}\epstev^{-2.5}\cmsqi\si\hbox{ster}^{-1}$, 
implying a number of $>1$~TeV background events 
$N_{bkg}\sim 10^{-4} (\theta /\hbox{deg})^2 t_2 {\rm km}^{-2}$
per angular resolution element over a burst duration of $10^2 t_2$~s. 
Thus, the neutrino signals from individual collapses at $z\sim 1$ would stand 
out with high statistical significance above the background.

The rate of $\gamma$-ray-detectable GRB events,
resulting from successful breakthroughs of jets beamed towards Earth,
is $\sim 10^3$/year \cite{fm95}. The predicted neutrino signal,
$\sim 0.2/{\rm km}^2$ muon events/burst at $\eps_{\nu_\mu}\simg 5 $ TeV
[Eq. (\ref{eq:N_mu})], would precede
by $t_{cross} \siml t_j = 10^2 t_2$~s the GRB trigger. Its detection 
would be a test of the collapsar hypothesis for long GRB, and the duration 
of the neutrino precursor would constrain the dimensions of, and jet advance
speed in, the stellar envelope. In this case the absence (or a much reduced 
duration) of $\nu$-precursors in short GRB would be a test of the view that 
long and short GRB arise from qualitatively different progenitors.

The signals from both emerging and choked-off jets in equation 
(\ref{eq:N_mu}) are calculated for ``mean'' events, i.e. for 
$E_{iso} \sim 10^{53}\E53$ erg, $\Gamma_j\sim 10^2\Gamma_2$ and $z\sim 1$. 
The total fireball jet energy and bulk Lorentz factors will, however, be 
distributed around these mean values, and fluctuations around the mean in 
these quantities and in the distance will lead to detections dominated by 
rare, energetic nearby events, as argued for GRB \cite{alvarez00}. Thus, one 
could expect a significant number of neutrino bursts (few/year) with 
$N_{\mu} \simg 10/{\rm km}^2$ muon events/burst.
% CH: ref B3 
These should be easily distinguishable above the atmospheric neutrino 
background at $z\sim 1$, whether coincident with observed GRB or not.

% CH3
The number of collapsars with Earth-pointing choked jets, while 
model-dependent, may be in excess of the number of successful collapsar 
GRB (where jet punch-through occurs and $\gamma$-rays are detected), 
$\simg 10^3$/year. If jet punch-through requires fast core rotation rate, 
e.g. from a massive star merger with a binary companion in a fraction 
of cases, choked collapsars could greatly exceed successful ones.
% ECH3
%If all were due to massive stellar collapses beamed towards Earth, and if
%we take $\Omega_j/4\pi \sim 10^{-2}$ as representative \cite{frail01},
%the total isotropic rate would be $\sim10^5$/year.
The number of massive collapses may, e.g., be comparable to that of type II 
supernovae (SNe II), $\sim 3\times 10^{-2}$/year/galaxy. 
For a galaxy density $n_G\sim 10^{-2}$/Mpc$^3$ within a Hubble radius $\sim 3$ 
Gpc ($z\sim 1$) the total rate of SNe II is ${\cal R}_{SN} \sim 1$ s$^{-1}$. 
If all SNe II lead to jets beamed into a solid angle $\Omega_j \sim 
10^{-2} 4\pi$ (e.g. \cite{frail01,cen98,wangwhee98}), the effective rate 
beamed towards Earth would be $\sim 10^5$/year. From equation (\ref{eq:fnu}), 
the $\nu\barnu$ energy flux per logarithmic energy interval from all 
directions would be 
$\epsilon_\nu^2\Phi_\nu \sim 10^{-7}{\rm GeV} / {\rm cm}^2{\rm s\,sr}$, 
where we divided by $\ln(\epsilon_{\nu,max}/\epsilon_{\nu,min})\sim 10$.
This signal is not far below the experimental upper bound on the diffuse 
neutrino flux recently established by the AMANDA experiment, 
$\epsilon_\nu^2\Phi_\nu \le 10^{-6}{\rm GeV/cm}^2{\rm s\,sr}$
\cite{amanda-bound}.
%atmospheric neutrino background 
%$4\pi \Phi_{bkg}\sim 10^{-6}\epstev^{-2.5} \cmsqi\si$. Thus, even if all
%core collapses resulted in relativistic jets, the maximum integrated 
%signal from the entire universe would not violate the diffuse background 
%constraints, and in fact could account for it at energies $\epstev \sim 1$.
An increase in neutrino telescope exposure will therefore allow in the
near future to put relevant constraints on the frequency of
$\gamma$-ray dark $\nu$-bursts from choked jets, and hence on
the frequency of core collapses at the poorly known high end of the stellar 
mass distribution. (Unfavorably aligned emerging jets
%, considered in a different cosmic ray context \cite{dermer00}, 
would also appear $\gamma$-ray dark, but would be neutrino-dark as well). 

The TeV neutrino signals preceding the $\gamma$-rays in GRB, as well as those 
from choked, $\gamma$-ray dark jets, would differ significantly from other
neutrino signals expected during and after the $\gamma$-ray phase of GRB.
In the latter, one expects $\simg 100$ TeV neutrinos \cite{wb97} from internal 
shocks well beyond the stellar envelope accelerating protons that interact
with MeV photons, or $\simg 10^{17}$ eV neutrinos \cite{wb00} from external 
shock protons interacting with UV photons even further out, and one also 
expects 2-5 GeV neutrinos \cite{bm00,mr00,dkk99} from inelastic nuclear 
collisions when neutrons decouple from protons. For the flat 
$\epsilon_\nu^2 \Phi_\nu$ neutrino spectra considered here, the number of 
precursor TeV neutrino events per burst is also larger by at least one order 
of magnitude, relative to those at $\simg 100$ TeV and at GeV expected during 
and after the $\gamma$-bright phase of GRB.
Such GRB precursor or $\gamma$-ray dark TeV neutrino burst signals may
% CH3
therefore be the likeliest targets for early detection with planned km$^3$ 
experiments \cite{halzen01}.

If, as suspected, the first generations of stars formed in the Universe are
much more massive than those forming now, then at redshifts $z\simg 5$ 
jet break-through collapses may be rare, and $\gamma$-ray dark choked jet 
collapses leading to TeV neutrinos may be much more common than today. 
Without an emerging jet, the X-ray, optical, IR and radio afterglows typical 
of GRB would also be absent. Such neutrino bursts,
which at $z\sim5$ produce a $\gtrsim1$~TeV fluence smaller by only a factor of
$\approx10$ compared to $z \sim 1$ bursts, would therefore be a prime tool for 
investigating both successful and failed GRB, as well as massive star formation 
at such high redshifts.

{\it Acknowledgements}: 
Partial support was received from BSF 9800343, Universities Planning \&
Budget Committee, Israel, NSF AST0098416 and NASA NAG5-9192.

\end{document}